\documentstyle[11pt,twoside,epsfig,jltp]{article}
\title{Phase diagram of the 1D Kondo lattice model}
\author{
I. P. McCulloch,
M. Gulacsi,\address{Department of Theoretical Physics,
Institute of Advanced Studies \\
The Australian National University,
Canberra, ACT 0200, Australia}
S. Caprara,\address{Dipertimento di Fisica - Universit\`{a}
di Roma ``La Sapienza'' \\
and Instituto Nazionale di Fisica della Materia, Unit\'{a} di Roma 1 \\
P. le A. Moro 2, I-00185 Roma, Italy} \\
A. Juozapavicius,${}^{\dagger}$
and A. Rosengren\address{Department of Theoretical Physics,
Royal Institute of Technology, S-100 44 Stockholm, Sweden}
}

\runninghead{I. P. McCulloch {\it et al.}}{Phase diagram of the 1D Kondo
lattice model}

\begin{document}
\newcommand{\beqa}{\begin{eqnarray}}
\newcommand{\eeqa}{\end{eqnarray}}
\newcommand{\beq}{\begin{equation}}
\newcommand{\eeq}{\end{equation}}
\newcommand{\dg}{\dagger}
\newcommand{\sig}{\sigma}
\newcommand{\vektor}[1]{\mbox{\boldmath $#1$}}
\begin{abstract}
We determine the boundary of the fully polarized ferromagnetic ground state
in the one dimensional Kondo lattice model at partial conduction
electron band filling by using a newly developed infinite size DMRG
method which conserves the total spin quantum number.  The obtained
paramagnetic to ferromagnetic phase boundary is below $J \approx 3.5$
for the whole range of band filling. By this we solve the controversy
in the phase diagram over the extent of the ferromagnetic region
close to half filling.

\end{abstract}

\maketitle

\vspace{0.3in}

The central problem posed by heavy fermion materials is to understand
the interaction between an array of localized moments (generally
f-electrons in lanthanide or actinide ions) and conduction electrons
(generally p- or d- band). The magnetic properties of the localized
moments and conduction electron interplay is well described
by the Kondo lattice model:
\beq
H \: = \: -t \: \sum_{\langle i,j \rangle, \: \alpha}
c^{\dg}_{i, \alpha} c^{}_{j, \alpha} \: + \:
J \: \sum_{j} \: {\bf S}_{c j} {\bf \cdot} {\bf S}_{j} \quad ,
\eeq
where $t > 0$ is the conduction electron
(or simply electron) hopping,
$\langle i,j \rangle$ denotes nearest neighbors,
${\bf S}_{j}$ are spin $1/2$ operators
for the localized spins,
and where ${\bf S}_{cj} = \frac{1}{2} \sum_{\alpha, \alpha '}
c^{\dg}_{j, \alpha} {\sig}_{\alpha, \alpha '} c^{}_{j, \alpha '}$
with $\sig$ the Pauli spin matrices and
$c^{}_{j, \alpha},c^{\dg}_{j, \alpha}$ the electron
site operators.

In the following, we concentrate on the region of the disordered
paramagnetic - ferromegnatic phase boundary which appears at
partial conduction band filling $n = N_{c}/L < 1$, where $N_{c}$
is the number of electrons, and $L$ the number of
localized spins, of the one dimensional
case. The transition has been identified
both analytically \cite{Honner}, and
in numerical simulations by a variety of methods
\cite{Troyer,Tsunetsugu,Moukouri95,Caprara}.
We focus on the $J > 0$ case, relevant to heavy-fermion systems,
in which the localized spins model $f$-electrons in lanthanide
or actinide ions.

The bulk of the numerical simulations
\cite{Troyer,Tsunetsugu,Moukouri95,Caprara}
are based on the DMRG approach and are devoted to
values of filling factor less than quarter filling.
Accordingly, our first focus is to extend the
region of the numerically available results up to
very close to half filling. However, our main focus is
to determine as accurately as possible the paramagnetic -
ferromagnetic phase boundary. For this we developed
a DMRG code which keeps track of the total spin. In this
way we can exactly determine the phase boundary where
the one dimensional Kondo lattice model becomes ferromagnetic.
These results will clarify the controversy \cite{RevModPhys}
over the paramagnetic - ferromagnetic phase boundary close
to half filling which has been a central issue for
much of the researh in this area for some years.

The density-matrix renormalization group (DMRG) method \cite{White}
has become a widespread method for the invetigation of (mostly) one
dimensional strongly correlated electron systems. Even though a lot
of new additions were made to the original method \cite{White} the
problem of keeping track of the total spin of the analysed model
was still a challenge. The problem emerges as a consequence of the
fact that the density matrix is not block diagonal
with respect to the total spin, except in the special case of $S = 0$.

In order to properly keep track of the ferromagnetic phase
appearing in the one dimensional Kopndo lattice model, we developed
a infinite DMRG method which conserves the total spin. This is the first
DMRG method that we are aware off to do this task. In the following
we give a brief description of the method, while a detailed analysis
will be published elsewhere \cite{Ian}.

Suppose the basis is partitioned into states of total spin, with the
basis vectors written as
\beq
\vert S, i \rangle, \quad i \: = \: 1,\: 2, \: \ldots, \: n_S \quad ,
\eeq
where $n_S$ is the number of basis states having total spin $S$.

Let the density matrix eigenstates be $\vert \rho_k \rangle$, with
corresponding  eigenvalues $\rho_k$. The density matrix in
the original basis is
\beq
\rho_{S_1, \: i; \: \: S_2, \: j} \: = \: \sum_k \: \rho_k \:
\langle S_1, \: i \: \vert \: \rho_k \rangle \:
\langle \rho_k \: \vert \: S_2, \: j \rangle \quad .
\eeq

\begin{center}
\begin{figure}
\epsfig{file=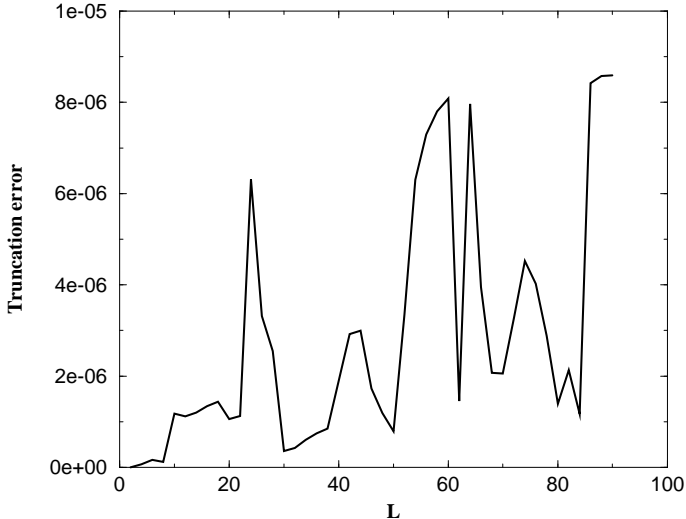,height=8cm}
\caption{The truncation error as a function of the length of
the chain for $n = 0.9$ and $J = 3.1$.}
\label{fig:fig1}
\end{figure}
\end{center}

The problem is to find the optimum basis state
\beq
\vert {\bf b} \rangle \: = \: \sum_{i=1}^{n_S} \: {\bf b}_i \:
\vert S, \: i \rangle \quad ,
\eeq
which is an eigenstate of total spin $S$.  After the best basis state
is found, the density matrix can be reduced in dimension and
the procedure applied iteratively until the best $m$ basis
states are found.

The measure of how good a basis state is, is given by the
probability, ${\cal P}(\vert {\bf b} \rangle)$, obtained from the
overlap with the original density matrix eigenstates,
\beq
{\cal P}(\vert {\bf b} \rangle) \: = \: \sum_k \: \rho_k \:
\vert \: \langle \rho_k \: \vert \: {\bf b} \rangle \: \vert^2 \quad .
\eeq

Maximizing ${\cal P}(\vert {\bf b} \rangle)$, subject
to the condition that $\vert {\bf b} \rangle$ is normalized, gives one
unique state, given by the highest weight eigenvector of the density
matrix with off-diagonal (with respect to total spin) elements removed.

\begin{center}
\begin{figure}
\epsfig{file=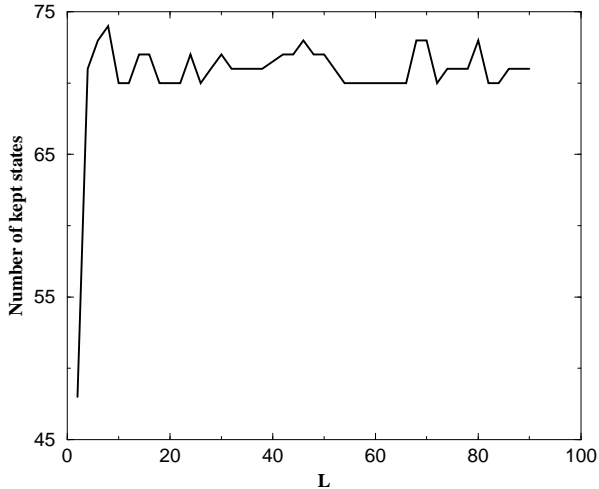,height=7.5cm}
\caption{The number of kept states, $m$ corresponding to Fig.\ \ref{fig:fig1}.}
\label{fig:fig2}
\end{figure}
\end{center}

Neglecting the off diagonal elements results in a tendency for the
spread of eigenvalues of the quasi-density matrix to be increased,
so it is necessary to keep more block states at each step.  However in every
case we have tried, this still leads to a significant overall
reduction in the number of states in the super block.
To exemplify, we present in Fig.\ \ref{fig:fig1} the truncation error.
As it can be seens it fluctuates to
accommodate all the states coorresponding to a given total spin value,
since it is necessary to keep all members of a spin multiplet to be able
to keep the Clebsch-Gordan transformation exact.
The corresponding number of kept states, $m$ is given in Fig.\ \ref{fig:fig2}.

We calculated the total spin of the system for $n = 0.5$,
$0.6$, $0.7$, $0.8$, $0.9$ and $0.95$ fillings. As the phase
transition is crossed, the total spin of the system changes from
0 to $(L - N_c) / 2$.  The fully polarized configuration is $N_c$ Kondo
singlets, and $L - N_c$ unscreened localized spins coupled ferromagnetically.
This type of configuration, seen to exist in the bosonization approach
\cite{Honner} also, signals the presence of spin polarons.

\begin{center}
\begin{figure}
\epsfig{file=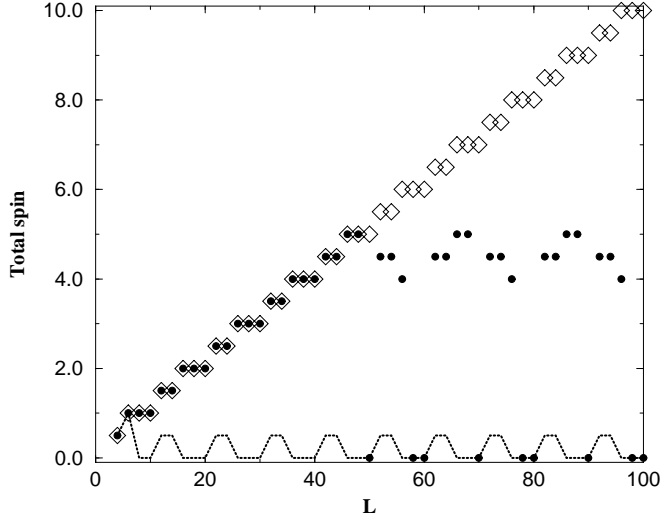,height=8cm}
\caption{The total spin as a function of the length of the chain for
$n = 0.8$. The dotted line corresponds to $J = 2.8$,
the filled circles to $J = 2.9$, while the open diamonds to $J = 3.0$.}
\label{fig:fig3}
\end{figure}
\end{center}

We have not yet determined from the numerical data the order of the
phase transition.  In Fig.\ \ref{fig:fig3} 
we present the case of $n = 0.8$. As it can be seen, the system is
paramagnetic up to $J = 2.8$.  The oscillations in the total spin are of
finite size origin.  It may be that immediately above the phase transition
there is a region where the total spin is not fully polarized.  This
would imply that the paramagnetic - ferromagnetic phase
transition in the one dimensional Kondo lattice model is not of first
order type, rather second order, or a quantum ordered-disordered phase
transition with variable exponents \cite{Honner}.  However, close to
the critical line, the total spin states become numerically degenerate 
so the results in this intermediate region are not reliable. Indeed,
we have not yet seen any fractionally polarized states where the
total spin can be determined unambiguously.

\newpage

In Fig.\ {\ref{fig:fig4} we present the phase diagram of the 1D Kondo lattice
model. We plotted only those cases which represent fully
polarized localized spins. Accordingly, these points
(denoted by stars in Fig.\ \ref{fig:fig4}) represent an upper bound on the
true thermodynamic phase transition.  As a comparison
we also plotted the previous numerical results (square
is DMRG data \cite{Moukouri95}, the diamond is the quantum Monte Carlo
data \cite{Troyer}, the open circles are the exact numerical diagonalization
data \cite{Tsunetsugu} and the filled circles are the infinite size DMRG
results \cite{Caprara}) and the phase transition curve obtained via
bosonization \cite{Honner}.

The most significant contribution of the present work is to establish the
existance of a ferromagnetic regime extending up to half filling, see
Fig.\ \ref{fig:fig4} for $n = 0.9$ and $n = 0.95$ filling factors.
For $n = 0.95$ the impurity spins are
already fully polarized above $J \approx 3.5$. This suggests that, contrary
to the ferromagnetic Kondo lattice model, in the present case ferromagnetism
will exist for any band filling.  For $J < 0$ the ferromagnetic phase only
assymptotically approaches half filling \cite{Honner,Yunoki}. The
reason for this is that a phase separated regime appears
\cite{Yunoki} between ferromagnetism and the insulating phase
at exactly half filling. Such a phase separated regime cannot appear for
the $J > 0$ Kondo lattice model and, as such, the present DMRG results
clearly confirms the predictions of the bosonization approach \cite{Honner}.

\newpage

\begin{center}
\begin{figure}
\epsfig{file=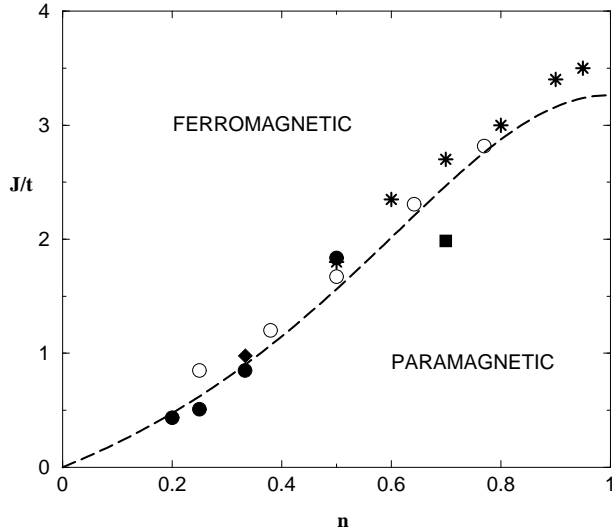,height=8cm}
\caption{The phase diagram of the 1D Kondo lattice model. The stars are the
results of the present work, they represent the first fully polarized
impurity spin states. For details, see text.}
\label{fig:fig4}
\end{figure}
\end{center}

Work in Australia was supported by the Australian Research
Council and DISR-IRAP. In Sweden by The G\"{o}ran Gustafsson
foundation, The Carl Trygger foundation and The Swedish
Natural Science Research Council.

\end{document}